\documentclass[lettersize,journal]{IEEEtran}
\usepackage{textgreek}
\usepackage{amsmath,amsfonts}
\usepackage{algorithm}
\usepackage{algpseudocode}
\usepackage{array}
\usepackage[caption=false,font=normalsize,labelfont=sf,textfont=sf]{subfig}
\usepackage{textcomp}
\usepackage{stfloats}
\usepackage{url}
\usepackage{verbatim}
\usepackage{graphicx}
\usepackage{cite}

\usepackage{amssymb}
\usepackage{nomencl}
\usepackage{etoolbox}

\usepackage{siunitx}

\begin{document}

\title{An Origami-Inspired Endoscopic Capsule with Tactile Perception for Early Tissue Anomaly Detection}

\author{Yukun~Ge,
        Rui~Zong,
        Xiaoshuai~Zhang,
        and~Thrishantha~Nanayakkara

\thanks{Y.Ge, R.Zong, X.Zhang, T.Nanayakkara are with Dyson School of Design Engineering, Imperial College London, SW7 2AZ London, UK. (e-mail: yukun.ge20@imperial.ac.uk).
*This work was supported in part by the Engineering and Physical Sciences Research Council (EPSRC) RoboPatient grant (EP/T00603X/1)}}

\markboth{}%
{Shell \MakeLowercase{\textit{et al.}}: A Sample Article Using IEEEtran.cls for IEEE Journals}


\maketitle

\begin{abstract}
Video Capsule Endoscopy (VCE) is currently one of the most effective methods for detecting intestinal diseases. However, it is challenging to detect early-stage small nodules with this method because they lack obvious color or shape features. In this letter, we present a new origami capsule endoscope to detect early small intestinal nodules using tactile sensing. Four soft tactile sensors made out of piezoresistive material feed four channels of phase-shifted data that are processed using a particle filter. The particle filter uses an importance assignment template designed using experimental data from five known sizes of nodules. Moreover, the proposed capsule can use shape changes to passively move forward or backward under peristalsis, enabling it to reach any position in the intestine for detection. Experimental results show that the proposed capsule can detect nodules of more than 3mm diameter with $100\%$ accuracy.
\end{abstract}


\section{Introduction}
Gastrointestinal cancer is becoming increasingly common, killing more than $3$ million people each year \cite{wang2023survey}. Diagnosing patients with gastrointestinal cancer is quite challenging because there are no typical symptoms in the early stages. Therefore, many patients are already in the middle or late stages of gastrointestinal cancer by the time they begin to feel uncomfortable, which leads to a high mortality rate \cite{wang2023survey}. Small intestinal cancer is a kind of fatal gastrointestinal cancer. The median age at diagnosis is $60-65$ years \cite{casali2022gastrointestinal}. In the United States, the overall 5-year survival rate is only $68.5\%$ \cite{SEER}. However, if patients receive an early diagnosis, prognosis, and treatment, the average 5-year survival rate can be increased to about $80\%$ \cite{giuliano2017gastric}. However, approximately one-third of patients are diagnosed with small intestinal cancer in a later stage \cite{aydin2016evaluation}. The evaluation and diagnostic process for small intestinal cancer includes imaging and endoscopic evaluation. Imaging methods such as computed tomography (CT scans) and X-rays find it difficult to detect the early stages of smaller tumors \cite{vuori1971primary}. Capsule endoscopy is currently the most commonly used method to detect cancerous masses in the small intestine \cite{soffer2020deep}. Because it is easier to enter the long and tortuous small intestine than a traditional tube endoscope. The current capsule endoscopes are based on Video Capsule Endoscopy (VCE). However, VCE relies on camera images of the bowel wall to detect relatively late symptoms such as abnormal shapes and colors. Capsule endoscopy has a high missed lesion rate and sometimes the lesions cannot be clearly recorded because of the lack of propulsion and control system \cite{zheng2012detection}. The latest advances involve the use of image processing and machine learning methods to help doctors diagnose tumors by predicting and highlighting their location with high accuracy \cite{ciuti2016frontiers}. However, these methods can only detect clearly visible tumors and detection accuracy drops dramatically for tumors $5$ mm in diameter or smaller \cite{kawano2015assessment}. 

In this letter, we present a type of origami capsule endoscope designed for the detection of early-stage tissue anomalies in the small intestine. The conceptual overview is shown in Fig.\ref{concept}. It consists of a capsule and an origami structure that wraps around the capsule. The capsule contains a processor, a battery, and a miniature motor for folding the origami structure. The origami structure is composed of four ring-shaped origami units, each made from a flexible piezoresistive material and covered with a flexible biocompatible material. The origami structure serves two functions: First, it controls the movement of the capsule. When the origami structure is unfolded, intestinal peristalsis propels it forward, and when it is in a folded state, peristalsis pushes it in the opposite direction. Secondly, it acts as a nodule detection sensor. When the intestinal wall compresses the origami structure, the resistance changes with the deformation of the structure. Signals can be analyzed using a particle filtering algorithm to detect potential nodules and estimate their size. 

The rest of this letter is organised as follows: Section \ref{design} presents the design approach and the functional features of the new origami capsule. The tactile sensor array and particle filter-based nodule detection are elaborated in Section \ref{sensing}. Section \ref{results} provides experimental validation results with indexes to quantify the effectiveness of the proposed approach for early detection of nodules in the intestine. Finally, section \ref{conclusions} provides conclusions.

\section{Design and Functional Features of the Origami Capsule}
\label{design}
\begin{figure}[tb]
\centering
\vspace{1mm}
\includegraphics[width=1\columnwidth]{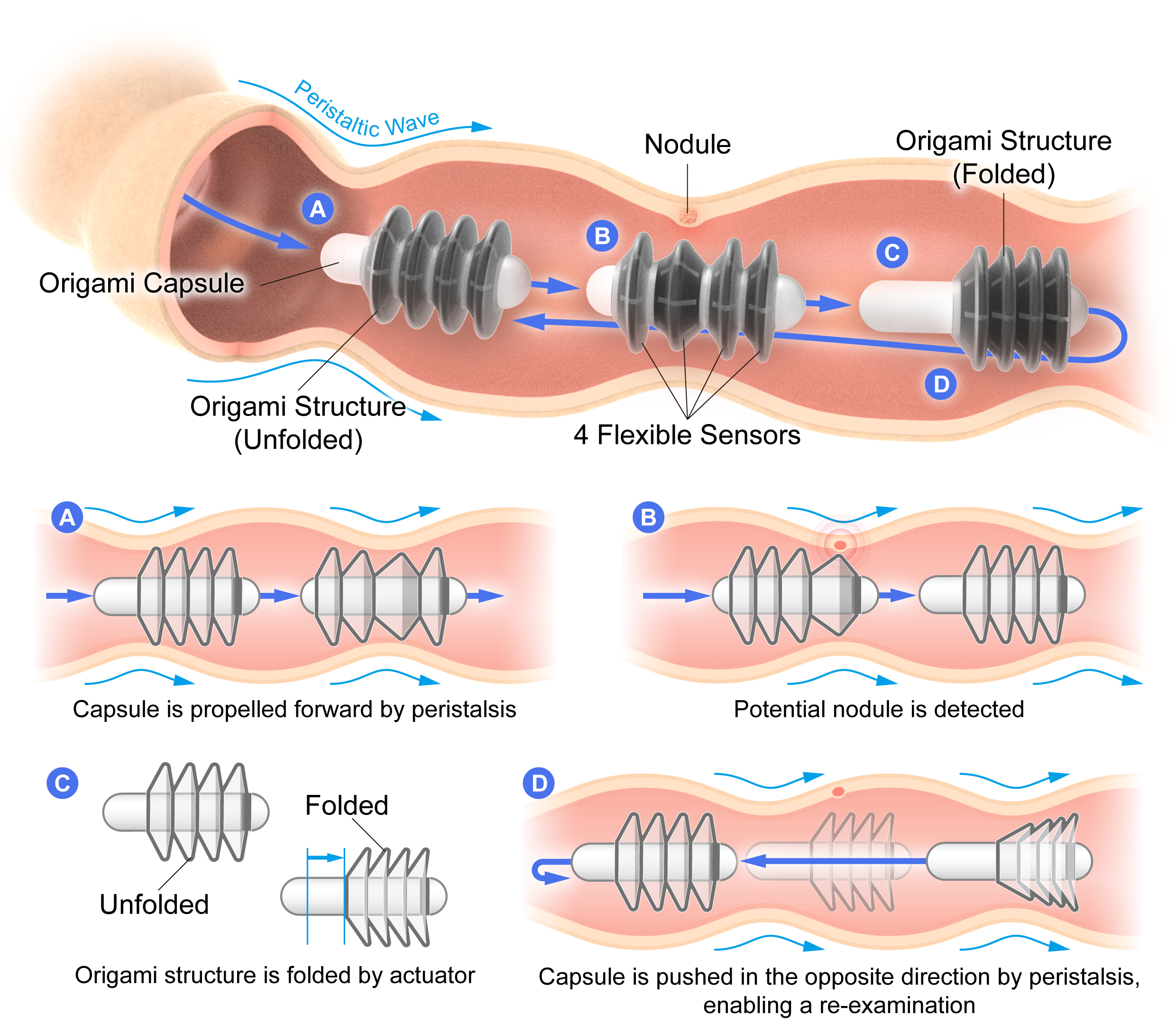}
\caption{Conceptual overview of origami capsule for early detection of intestinal tissue anomaly. (A) When the origami structure of the capsule is unfolded, it is continuously propelled forward by the peristalsis of the intestine. (B) The origami structure also serves as a flexible pressure sensor. As the capsule moves forward, intestinal pressure data is continuously collected. Potential nodule signals can be analyzed using an algorithm. (C) Once a signal resembling a nodule is detected, the origami structure will be folded up by an actuator inside the capsule. (D) The folded origami structure will be pushed backward by intestinal peristalsis, returning to the location of the abnormal signal for retesting to improve diagnostic accuracy.}
\label{concept}
\end{figure}

The process of the origami capsule for nodule detection is as follows: With the origami structure in an unfolded state, the capsule is propelled forward by intestinal peristalsis as shown in Fig.\ref{concept}(A). During the forward movement of the capsule, sensors continuously collect and process signals as shown in Fig.\ref{concept}(B). When a signal resembling a nodule is detected, the size of the nodule is analyzed. If necessary, the capsule can return to the suspected nodule location for resampling. The origami structure can be folded by a miniature motor as shown in Fig.\ref{concept}(C). The capsule is then pushed backward by intestinal peristalsis to the previous position as shown in Fig.\ref{concept}(D). Afterward, the origami structure unfolds again, moving forward to collect more data.

\subsection{Origami Structure Design}


The inspiration for the origami structure comes from wheat ears. By repeatedly squeezing and releasing a wheat ear in the hand, it will be found to move in the direction opposite to that of the awns. In the previous research, we described the initial version of the origami structure, whose unfolded form would be pushed backward by the intestine. Once completely contracted into a rigid body, it could be propelled forward by peristalsis \cite{9730084}. 

Since we analyze and record the deformation caused by the intestinal squeezing of the origami structure to detect nodules, the origami structure needs to maintain elasticity and frequent contact with the intestine as it moves forward. In this study, the origami structure was modified to the new geometric shape shown in Fig.\ref{manufacturing}(A). The origami structure is composed of four substructures connected in series. The right side is the head of the origami structure, and the left side is the tail. Each substructure is a symmetrical hexagon $ABCA'B'C'$, aligned along the black dotted line shown in the diagram. The line segment $BC$ extends outward a short distance to point $D$. Since the origami structure is made of elastic material, the function of the hexagon is to allow the origami structure to act like a spring, stretching and contracting under the periodic squeezing of the intestinal peristaltic waves. The extended segment $CD$ acts like the awn of a wheat ear, serving to constrain the direction of movement of the origami structure. The  ``awn'' $D$ features a gentle chamfer with a radius $r$ of $1$ mm to prevent damage to the intestinal wall. The width $w$ of the substructure is $20$ mm, which is less than the width of the resting intestine. The vertical distance $h_1$ from the symmetry axis to $A$ is $2$ mm, the vertical distance $h_2$ from $A$ to $C$ is $5$ mm, and the vertical distance $h_3$ from $C$ to $D$ is $3$ mm. The angle $\beta$ between $CA$ and $CB$ is $45^\circ$. The angle $\alpha$ between line segment $BD$ and the vertical direction is $5^\circ$ counterclockwise, so the ``awn'' all point to the left, causing the origami structure to move to the right when compressed.

\subsection{Origami Structure Fabrication}

\begin{figure}[tb]
\centering
\vspace{1mm}
\includegraphics[width=1\columnwidth]{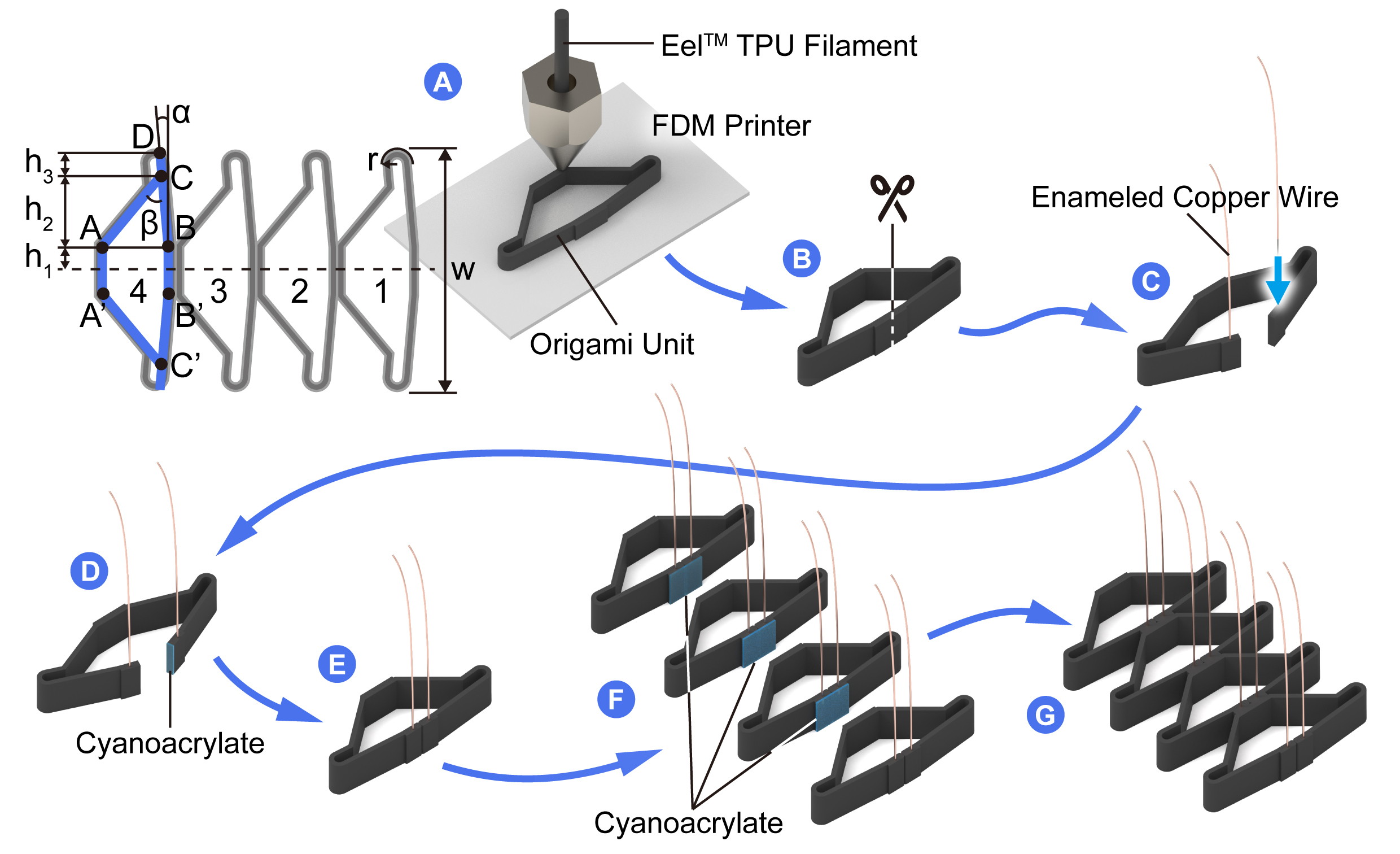}
\caption{Origami structure fabrication method. (A) $4$ Independent origami substructures were printed using an FDM printer. The filament was conductive Eel\textsuperscript{TM} TPU. (B) The origami substructure was cut along the dotted line. (C) An enamelled copper wire was inserted into each cut end of the substructure. (D)-(E) The cut part of the substructure was reattached using cyanoacrylate adhesive. (F)-(G) 4 substructures were glued together to form a complete origami structure.}
\label{manufacturing}
\end{figure}

The origami structure was produced through the use of FDM (fused deposition modelling) printing technology, employing a Prusa i3 MMU2S printer. The material used was NinjaTek Eel\textsuperscript{TM} TPU (Thermoplastic polyurethanes) 3D printing filament ($1.75$ mm diameter, Shore Hardness $90$ A, containing $18$ wt\% carbon black) \cite{EELDatasheet}. The electrical resistance of flexible sensors made from NinjaTek Eel\textsuperscript{TM} TPU changes with shape deformation. In recent years, some researchers have utilized this piezoresistive material to design wearable electronic fabrics for monitoring limb movements \cite{glogowsky2023influence}\cite{goncu20233d}\cite{heracleous2022scalable}. However, reports on utilizing this material for creating origami structures or tactile sensors are still scarce. 

The fabrication process is as follows. First, as shown in Fig.\ref{manufacturing}(A), four origami substructures were printed independently. The height of the substructures was $5$ mm, and the thickness was $0.5$ mm. Then, the origami substructure was cut along the dotted line in Fig.\ref{manufacturing}(B). As shown in Fig.\ref{manufacturing}(C), to measure the resistance of each substructure. Wires needed to be connected to both ends of each substructure. The insulating paint on the tips of $0.1$ mm diameter enamelled copper wires was removed by heating. Wires were then soldered into both ends of the substructure (the positions of the blue arrows in Fig.\ref{manufacturing}(C)) through an approximately $400\si{\celsius}$ soldering iron. Reattach the cut surface of the substructure with cyanoacrylate adhesive, ensuring that the contact surfaces were fully coated with the adhesive to prevent short-circuiting and ensure accurate resistance readings, as shown in Fig.\ref{manufacturing}(D) and Fig.\ref{manufacturing}(E). Each substructure was glued together with cyanoacrylate adhesive to form a complete origami structure, as shown in Fig.\ref{manufacturing}(F) and Fig.\ref{manufacturing}(G). The area where the units are connected needs to be fully coated with adhesive to ensure that the units are insulated and prevent signal interference.

\subsection{Small Intestine Simulator}

\begin{figure}[tb]
\centering
\vspace{1mm}
\includegraphics[width=0.8\columnwidth]{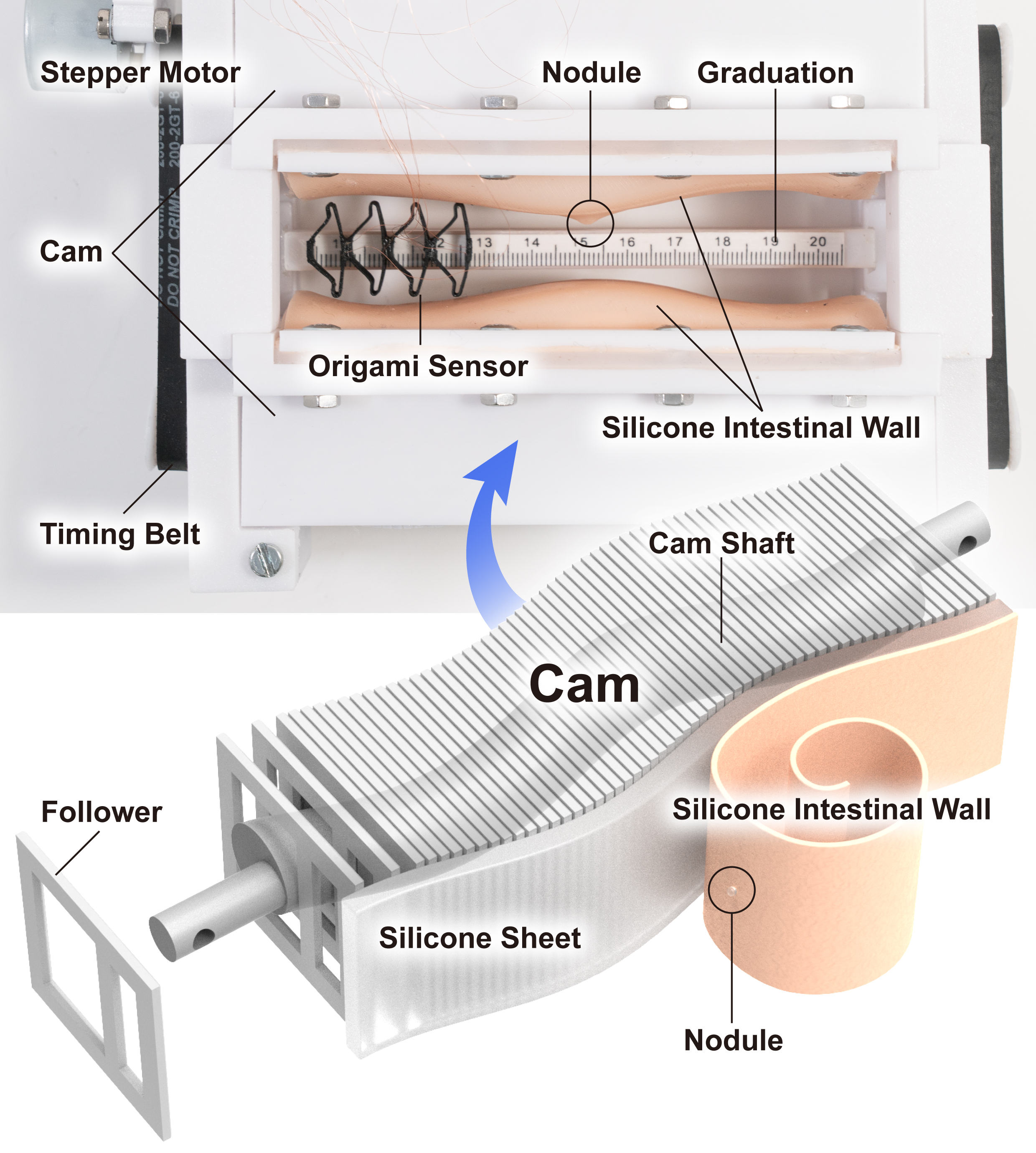}
\caption{Overall experiment set-up. We tested the origami sensor in a two-dimensional intestinal simulator. Two silicone intestinal walls were made of Ecoflex\textsuperscript{TM} 00-10. A sphere made of rubber was placed under the intestinal wall to simulate a nodule. Peristalsis was generated by squeezing the intestinal walls with cam mechanisms. The silicone sheet placed between the cam and the intestinal wall was used to make the peristalsis more elastic. The speed of peristalsis can be adjusted by controlling a 28BYJ-48 stepper motor rotation through Arduino UNO. Arduino Mega 2650 was used to record the resistance of each sensor unit on the origami sensor.}
\label{setup}
\end{figure}

We designed a set of experiments to test the origami structure's ability to detect nodules.  A two-dimensional small intestinal simulator was designed to simulate simplified intestinal environments such as peristalsis and masses.

Fig.\ref{setup} shows the small intestine simulator set-up. Two parallel intestinal walls, each $1$ mm thick, were made of Ecoflex\textsuperscript{TM} 00-10 because it had similar mechanical properties to intestinal tissue, which the moduli is $50$ kPa \cite{vaicekauskaite2020mapping}. nodules were simulated using spheres of $1$ mm, $2$ mm, $3$ mm, $4$ mm, and $5$ mm, made from PT Flex 60 rubber \cite{PTflexDatasheet}. For each test, only one nodule was used, and it was placed underneath the intestinal wall. There were a pair of symmetrical cam mechanisms on the outside of the silicone sheets to simulate intestinal peristalsis. The cam mechanism consisted of a row of $2$ mm thick hollow rectangular followers and a spiral cam shaft running through them. (Thinner followers can produce a smoother and more continuous curve of intestinal peristalsis.) The upper and lower sides of the followers were clamped by the slides, so they could only slide in the horizontal direction. A 28BYJ-48 Stepper Motor was installed on the end of one of the cams to rotate. Timing belts were installed at both ends of the cam shafts to ensure that their rotations were synchronous and symmetrical. When the motor rotated, the cams compressed the silicone sheets to form continuous sinusoidal waves with a wavelength of $80$ mm, similar to the actual intestinal peristalsis \cite{SINNOTT2017143}\cite{srivastava2007effects}. Since there are few studies describing the peristaltic amplitude, we set the amplitude to $3.5$ mm in this experimental set-up. The lumen diameter of the small intestine is about $20$ mm in a resting state \cite{cronin2010normal}. Since the intestinal walls were sinusoidal shapes in the setup, we considered the distance between the sinusoidal baselines of the two silicone walls as the diameter of the small intestine, which is $20$ mm. Since peristalsis is caused by the contraction of circular muscles and has a certain elasticity, we placed a $3$ mm thick Ecoflex 00-10 silicone sheet between the rigid cam and the intestinal wall to make the peristalsis more realistic  \cite{huizinga1999gastrointestinal}. Between two sections of the intestinal wall, there is a platform marked with measurements, designed to hold the origami structure. The height at which the origami structure is positioned allows it to come into contact with the nodule, and the measurements assist us in observing the distance the origami structure moves when necessary.

\section{Tactile Sensing and Signal Processing}
\label{sensing}
\subsection{Data Collection}

In the experiment, sensor readings from the origami structure were collected using an Arduino MEGA 2560 R3, with each substructure connected in series to a $12$ k\textOmega{}  resistor (which was similar to the resistance of the substructure itself). The Arduino recorded the voltage values across each substructure at $10$ Hz. Before each experiment began, the origami structure was placed on the left side of the intestine, as shown in the current position in Fig.\ref{setup}. Then, the phantom intestine started to exhibit peristaltic movements. The direction of peristalsis was from left to right, with a wave propagation speed of $25$ mm/s \cite{avvari2015bio}. The recording ceased when the peristalsis pushed the origami structure to the far right.  

The experiment was conducted separately for the following six conditions: no nodule and a nodule with diameters of $1$ mm, $2$ mm, $3$ mm, $4$ mm, and $5$ mm. The selection of a maximum diameter of $5$ mm for the nodules was motivated by the current limitations of capsule endoscopy, which struggles to detect nodules of this size. In the research conducted by Gilabert et al, it was found that doctors reviewing images returned by an endoscope might even miss polyps with diameters of $6$ or $7$ mm \cite{gilabert2022artificial}.

\subsection{Data Pre-processing}

\begin{figure}[tb]
\centering
\vspace{1mm}
\includegraphics[width=1\columnwidth]{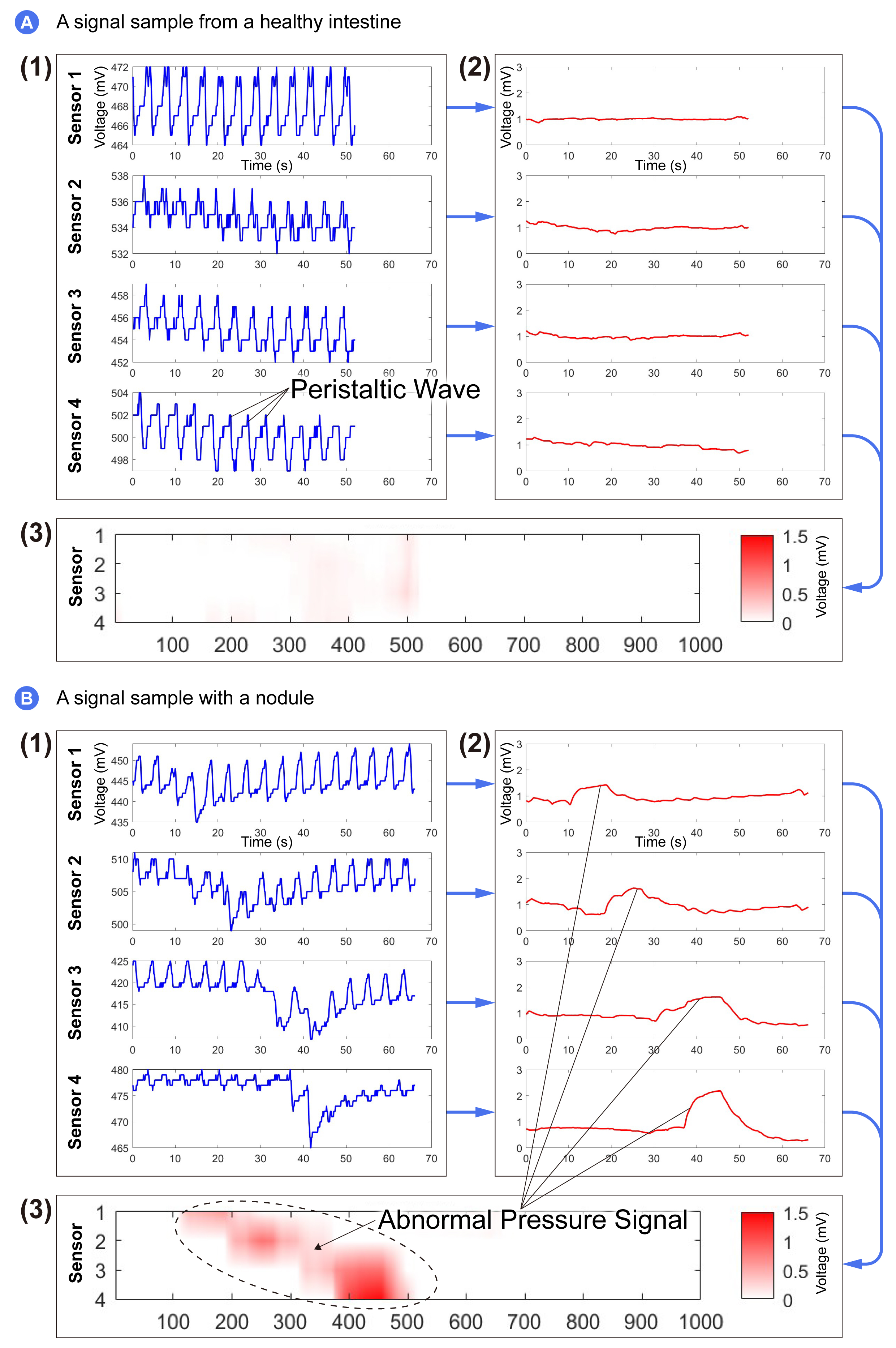}
\caption{The data pre-processing process. (A) In the case of a healthy intestine without a nodule: (A)(1) Original signal from a healthy intestine without a nodule, (A)(2) RMS envelope (window size \( W = 84 \)) result of the signal, (A)(3) Matrix formed by merging the signals from four sensors. (B) In the case of an intestine with a nodule: (B)(1) Original signal from an intestine with a nodule, (B)(2) RMS envelope result showing that the sensors sequentially detect a sharp increase in abnormal pressure, (B)(3) A matrix where a distinctive pattern is visible, indicating the presence of a nodule. }
\label{preprocessing}
\end{figure}

We performed pre-processing on each set of collected data, as shown in Fig.\ref{preprocessing}. Each dataset contains signals from four sensors, with the sensor sequence $i$ following the order from right to left as shown in Fig.\ref{manufacturing}(A), with labels $1$, $2$, $3$, and $4$. Fig.\ref{preprocessing}(A)(1) displays data from a healthy intestine without a nodule, showing the pressure signal resulting from intestinal peristalsis. Fig.\ref{preprocessing}(B)(1) shows data with a nodule, where the nodule's compression of the sensors causes signal changes. We extract the characteristic signal of the nodule through the following steps.

First, each sensor's data is normalized to have a mean of $0$ and a standard deviation of $1$. The purpose of this step is to eliminate the threshold differences in readings from the flexible sensors, which can have individual variations and can be challenging to calibrate, as follows:
\begin{equation} \label {normalization} 
\hat{\mathbf{v}} = \frac{ \mathbf{v} - \mu}{\sigma},
\end{equation} 
where $\hat{\mathbf{v}}$ is the normalized data, $\mathbf{v}$ is the original data of one sensor, $\mu$ is the mean, and $\sigma$ is the standard deviation of data.

Next, the Root Mean Square (RMS) envelope of $\hat{\mathbf{v}}$ is calculated \cite{seryasat2010multi}. This step helps to remove intestinal peristaltic wave signals and reveals potential pressure changes on the sensors due to nodule compression. From Fig.\ref{preprocessing}(B)(2), it can be observed that the pressure signal from the nodule has been extracted from the original signal containing peristaltic signals, as follows:
\begin{equation} \label {RMS_envelope} 
\tilde{\mathbf{v}}(k) = \sqrt{\frac{1}{W'(k)} \sum_{j = \max(1, k - \frac{W}{2})}^{\min(K, k + \frac{W}{2})} \hat{\mathbf{v}}(k)^2},
\end{equation} 
where $\tilde{\mathbf{v}}$ represents the RMS envelope result of $\hat{\mathbf{v}}$, \( k \in \{1, 2, \ldots, K\} \) is the index of the current sample in the vector, $K$ is the total number of samples in the vector, \( W = 84 \) is the window size, and $W'(k)$ is the actual window size used for calculation, it dynamically changes based on the position $k$, and is determined by the following piecewise function:
\begin{equation} \label {W'} 
W'(k) = 
\begin{cases} 
k + \frac{W}{2}, & \text{if } k < \frac{W}{2} , \\[1em]
W, & \text{if } \frac{W}{2} \leq k \leq K - \frac{W}{2} ,\\[1em]
K - k + \frac{W}{2}, & \text{if } k > K - \frac{W}{2} ,
\end{cases}
\end{equation}
for the middle portion of the signal (\(\frac{W}{2} \leq k \leq K - \frac{W}{2}\)), the window size remains \(W\). At the beginning boundary of the signal (\(k < \frac{W}{2}\)), the window size increases from $1$ to $W$, reflecting the actual available data points. At the ending boundary of the signal (\( k > K - \frac{W}{2}\)), the window size decreases from $W$ to $1$, similarly reflecting the actual available data points.

The origami structure passes continuously through the nodule, and the sensors are discrete. Therefore, when the nodule is between two sensors, both of these sensors will feel it at varying levels. However, we can still treat the data from the sensors as ``continuous'' for analysis purposes. Thus, the data from the four sensors are merged into a single matrix for analysis. The envelope data from the sensors are constructed into a matrix as follows:
\begin{equation} \label {constructed_matrix} 
\tilde{\mathbf{V}} = \begin{bmatrix}
\tilde{\mathbf{v}}_1,~
\tilde{\mathbf{v}}_2,~
\ldots,~
\tilde{\mathbf{v}}_S
\end{bmatrix}^T,
\end{equation}
where $\tilde{\mathbf{V}}$ is the merged matrix, \( S=4 \) is the total number of sensors.

To highlight the abnormal pressure features, data below a certain threshold are ignored. Here, the ReLU (rectified linear unit) activation function is used, which involves subtracting a threshold from each element in the data and setting elements smaller than $0$ to $0$ \cite{kessler2017application}, as follows:
\begin{equation} \label {ReLU}
\mathbf{V}_{\text{ReLU}}(i, k) = \max \left( \tilde{\mathbf{V}}(i, k) - c, 0 \right),
\end{equation}
where \(\mathbf{V}_{\text{ReLU}}\) is the matrix after ReLU processing, \( c=1 \) is the subtracted threshold.

The total duration of data collection varies each time, meaning that the total sample count $K$ is different. For the convenience of subsequent calculations, the \(\mathbf{V}_{\text{ReLU}}\) matrix is extended to a uniform length, as follows:
\begin{equation} \label {extended} 
\mathbf{V} = 
\begin{cases} 
\mathbf{V}_{\text{ReLU}}(i, k), & \text{for } 1 \leq i \leq S \text{ and } 1 \leq k \leq K, \\[1em]
0, & \text{for } 1 \leq i \leq S \text{ and } K < k \leq L,
\end{cases}
\end{equation}
where $\mathbf{V}$ is the extended matrix with dimensions of $S \times L$, \( L=1000 \) is the number of columns in the extended matrix.

Fig.\ref{preprocessing}(A)(3) and Fig.\ref{preprocessing}(B)(3) respectively show the signal matrices without a nodule and with a nodule. The ``contour'' syntax in MATLAB was used here to plot the matrices \cite{ContourMatlab}. Although the matrix has only $S$ rows, the ``contour'' syntax visually fills in the gaps between each row for ease of observation, without altering the original matrix. We can easily observe clear patterns in the data with nodules, but there are no patterns in the data without nodules.

The above is the data pre-processing process. In a single experiment, time-based voltage data $\mathbf{v}$ from four sensors are normalized, RMS envelope is calculated, merged into a matrix, undergoes ReLU processing, and extended to create the matrix $\mathbf{V}$ for subsequent analysis.

\subsection{Template Generation}

\begin{figure}[tb]
\centering
\includegraphics[width=1\columnwidth]{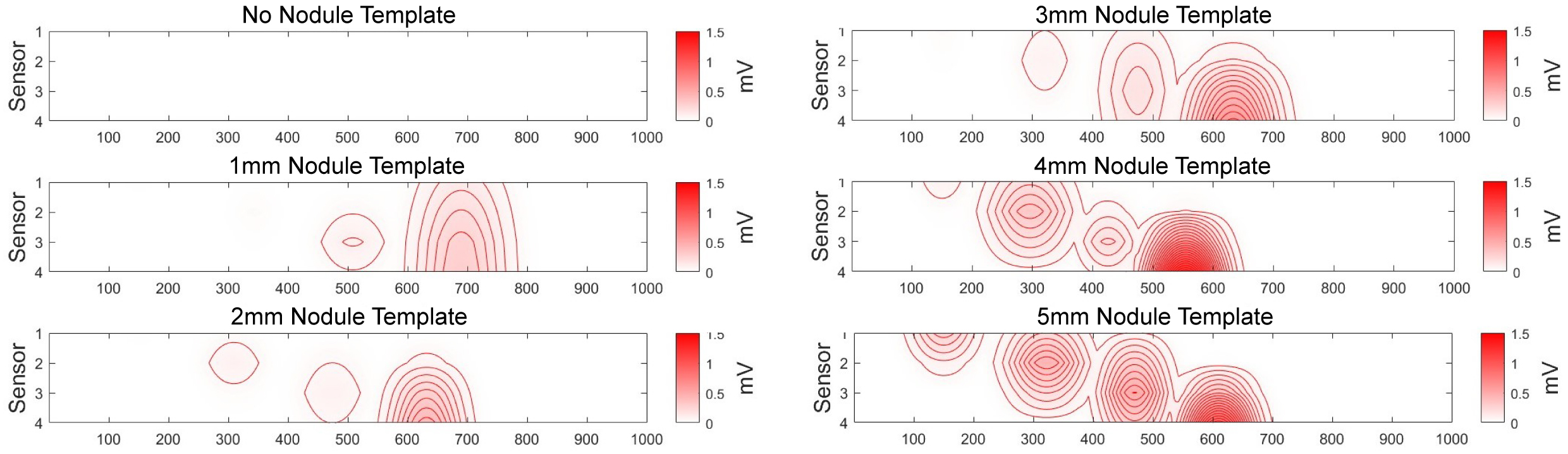}
\caption{The templates include the template without nodules ($\mathbf{T}^0$) and the templates for nodules of sizes $1$ mm to $5$ mm ($\mathbf{T}^1$ to $\mathbf{T}^5$). The visualization of templates includes contour lines to facilitate observing their differences.}
\label{general_template}
\end{figure}

Through experiments, it was discovered that nodules of different sizes leave different patterns. It is hypothesized that templates for nodules of different sizes can be generated using particle filtering methods. The test data will be compared with the templates to detect the presence of nodules and assess their sizes. The template generation method is as follows: 1. Prepare a matrix of specified nodule size; 2. Find the particle from a randomly generated set of particles that is most similar to the data; 3. Resample and update the particles to make them more similar; 4. Average the particles generated from several data sets with the same nodule size to form a template. The methodology of particle filtering is based on the research of Elfring et al \cite{elfring2021particle}.

Firstly, we pack several pre-processed experimental data sets with nodules ranging from $1$ mm to $5$ mm into a set, used for generating templates for each nodule size, as follows:
\begin{equation} \label {expdataset}
\mathcal{F} = \left\{ \mathbf{F}^\zeta \mid \zeta = (b, q) \right\},
\end{equation}
where $\mathcal{F}$ is the collection of all experimental data used for generating templates, $\mathbf{F}^\zeta$ is the pre-processed experimental data matrix, \( b \in \{1,2,\ldots,5\} \) is the size of the nodule, \( q \in \{1,2,\ldots,Q\} \) is the index of the data sample for nodule size $b$, \( Q = 20 \) is the total number of the data.

For all experimental data $\mathbf{F}^\zeta$ in $\mathcal{F}$, we generate a particle similar to that data.
According to Equations (\ref{generateP}) to (\ref{G}) described later, we can generate a set of initial particles and pack these particles with their corresponding parameters into one set, as follows:
\begin{equation} \label{particleSet}
\mathcal{P}^\zeta = \left\{ (\mathbf{P}_{\varphi}^\zeta, \Theta_{\varphi}^\zeta) \mid  \varphi = (n, m) \right\},
\end{equation}
where $\mathcal{P}^\zeta$ is the set of particles, $\mathbf{P}_{\varphi}^\zeta$ is a particle, $ \Theta_{\varphi}^\zeta $ is the set of parameters for that particle, \( n \in \{ 1,2, \ldots, N \} \) is the index of the particle, $N=2000$ is the total number of particles, \(m = 0\) is the number of iterations for the particle (since it is the initialization stage, the value is set to $0$). $ \Theta_{\varphi}^\zeta $ is constructed as follows:
\begin{align} \label {particleparameters}
\begin{split}
\Theta_{\varphi}^\zeta = \{ &A_{\varphi,s}^\zeta , \boldsymbol{\mu}_{\varphi,s}^\zeta , \boldsymbol{\Sigma}_{\varphi,s}^\zeta \mid s = 1, 2, \ldots, S \},
\end{split}
\end{align}
in Equations (\ref{generateP}) to (\ref{G}), the meanings of all variables in the parameter set are explained.

Observing from Fig.\ref{preprocessing}(B)(3) that the pattern of the experiment data with nodule roughly consists of four elliptical shapes connected from the top left to the bottom right. This is because the nodule sequentially presses on four sensors, leaving four impressions. Therefore, we can generate a particle using Gaussian surfaces arranged diagonally from the top left to the bottom right to simulate the pattern of experimental data, as follows:
\begin{equation} \label {generateP}
    \mathbf{P}_{\varphi}^\zeta(i, j) = \max \left\{ \mathbf{G}_{\varphi,s}^\zeta(i, j) \mid s = 1, 2, \ldots, S \right\},
\end{equation}
where \( i \in \{1, 2, \ldots, S\} \), \(j \in \{1, 2, \ldots, L\} \), $\mathbf{G}_{\varphi,s}^\zeta$ represents the $s$th Gaussian surface, $s$ is also the index of the sensors. The equation for generating the Gaussian surfaces $\mathbf{G}_{\varphi,s}^\zeta$ is as follows:
\begin{equation} \label {G}
    \mathbf{G}_{\varphi,s}^\zeta(i, j) = A_{\varphi,s}^\zeta \mathcal{N}\left( \boldsymbol{\mu}_{\varphi,s}^\zeta, \boldsymbol{\Sigma}_{\varphi,s}^\zeta \right), 
\end{equation}
where \(A_{\varphi,s}^\zeta \in (0, 1.5] \) is the amplitude, the maximum value of $1.5$ for the range is derived from the maximum value of the collected data, \(\mathcal{N}(\cdot, \cdot)\) represents the multivariate Gaussian distribution, $\boldsymbol{\mu}_{\varphi,s}^\zeta$ is the mean vector, and $\boldsymbol{\Sigma}_{\varphi,s}^\zeta$ is the covariance matrix. The structure of $\boldsymbol{\mu}_{\varphi,s}^\zeta$ is as follows:
\begin{equation} \label {mu}
\boldsymbol{\mu}_{\varphi,s}^\zeta = 
\begin{bmatrix}
  \mu_{\varphi,s}^{\text{ver},\zeta}  & \mu_{\varphi,s}^{\text{hor},\zeta}
\end{bmatrix}^T, 
\end{equation}
where \(\mu_{\varphi,s}^{\text{ver},\zeta} = s\) represents the mean along the vertical ($i$) direction in the matrix, because each row ($i$) of the matrix is composed of readings from sensor $s$ as in Equation (\ref{constructed_matrix}), $\mu_{\varphi,s}^{\text{hor},\zeta}$ is the mean in the horizontal ($j$) direction, which is the time step at which the peak value appears for each sensor. The expression for $\mu_{\varphi,s}^{\text{hor},\zeta}$ is as follows:
\begin{align} \label {deltaMu}
\begin{split}
\mu_{\varphi,s}^{\text{hor},\zeta} &= 
\begin{cases} 
 u, & \text{if } s = 1, \\[1em]
 \mu_{\varphi,{s-1}}^{\text{hor},\zeta} + \Delta \mu_{\varphi,s}^{\text{hor},\zeta}, & \text{if }  s > 1,
\end{cases} 
\end{split}
\end{align}
where  \(u=150\) represents the mean in the $j$ direction for the first Gaussian surface, \(\Delta \mu_{\varphi,s}^{\text{hor},\zeta}\in \{100, 101, \ldots, 250\}\) represents the increment in the mean in the $j$ direction for the next Gaussian surface. The gradual increase in $\mu_{\varphi,s}^{\text{hor},\zeta}$ is due to the capsules moving forward to detect nodules, whereby the nodules sequentially compress onto each sensor. The range of $\Delta \mu_{\varphi,s}^{\text{hor},\zeta}$ is selected based on the characteristics of the collected data. The structure of $\boldsymbol{\Sigma}_{\varphi,s}^\zeta$ is as follows:
\begin{equation} \label {sigma}
 \boldsymbol{\Sigma}_{\varphi,s}^\zeta = 
  \begin{bmatrix}
  \left( \sigma_{\varphi,s}^{\text{ver},\zeta} \right)^2&0 \\[1em]
  0&\left( \sigma_{\varphi,s}^{\text{hor},\zeta} \right)^2
 \end{bmatrix},
\end{equation}
where \(\sigma_{\varphi,s}^{\text{ver},\zeta} \in (0, 1] \) is the standard deviation in the $i$ direction, the maximum value of $1$ is because when the nodule is between two sensors, both sensors may detect a signal, causing the Gaussian surface of a single sensor to extend to the adjacent sensors (adjacent rows in the matrix). \(\sigma_{\varphi,s}^{\text{hor},\zeta} \in \{1, 2, \ldots, 60\} \) represents the standard deviation in the $j$ direction, which can be understood as the duration of time steps during which each sensor detects the nodule signal. The range of $\sigma_{\varphi,s}^{\text{hor},\zeta}$ is selected based on the characteristics of the collected data. 

For the next step, We find a particle in the set $\mathcal{P}^\zeta$ that is most similar to the experimental data $\mathbf{F}^\zeta$. The criterion for comparison is to search for the particle with the smallest RMSE value when subtracted from the experimental data. 

Since the occurrence timing of the nodule signal in the experimental data is uncertain, to compare with particles, we need to slide the matrix of particles or templates in the $j$ direction of the experimental data matrix. This allows us to find the position where the difference between the two matrices is minimal for comparison. The sliding starts with the last column of $\mathbf{P}_{\varphi}^\zeta$ aligned with the first column of $\mathbf{F}^\zeta$. Then, $\mathbf{P}_{\varphi}^\zeta$ starts sliding to the right until the first column of $\mathbf{P}_{\varphi}^\zeta$ aligns with the last column of $\mathbf{F}^\zeta$, and the sliding ends. To maintain a constant matrix size for the overlapping parts during sliding, and to ensure that the overlapping part contains both complete matrices $\mathbf{F}^\zeta$ and $\mathbf{P}_{\varphi}^\zeta$ for easy comparison of differences after each slide, we extend both sides of $\mathbf{F}^\zeta$ by adding zero matrices of size $S \times L$, as follows:
\begin{equation} \label {F'}
\mathbf{F'}^{\zeta} = 
\begin{bmatrix}
0_{S \times L} & \mathbf{F}^\zeta & 0_{S \times L}
\end{bmatrix} ,
\end{equation}
where $\mathbf{F'}^{\zeta}$ is the extended version of $\mathbf{F}^\zeta$. Then, we extend the right side of $\mathbf{P}_{\varphi}^\zeta$ with a zero matrix of size $S \times 2L$, as follows:
\begin{equation} \label {P'}
\mathbf{P'}_{\varphi}^{\zeta} = 
\begin{bmatrix}
\mathbf{P}_{\varphi}^\zeta & 0_{S \times 2L}
\end{bmatrix} ,
\end{equation}
where $\mathbf{P'}^{\zeta}$ is the extended version of $\mathbf{P}^\zeta$. Then, we multiply $\mathbf{P'}_{\varphi}^{\zeta}$ by a right shift matrix, where all elements in $\mathbf{P'}_{\varphi}^{\zeta}$ are shifted one position to the right, achieving the effect of ``sliding''. Referring to the case of cyclic shift matrix (CSM) by Zhu et al.\cite{CyclicShiftMatrix}, here we use a non-cyclic shift matrix, meaning that the vacant elements after shifting are filled with a zero matrix. The right-shift matrix $\mathbf{R}$ is a specially constructed matrix, with the following form:
\begin{equation} \label {shiftMatrix}
\mathbf{R} = 
\begin{bmatrix}
0 & 1 & 0 & \cdots & 0 \\
0 & 0 & 1 & \ddots & \vdots \\
\vdots & \vdots & 0 & \ddots & 0 \\
0 & 0 & \vdots & \ddots & 1 \\
0 & 0 & 0 & \cdots & 0
\end{bmatrix}_{3L \times 3L} .
\end{equation} 
Each time the matrix slides one position, calculate the Root Mean Square Error (RMSE) between  $\mathbf{F'}^{\zeta}$ and  $\mathbf{P'}_{\varphi}^{\zeta}$. The formula is as follows:
\begin{equation} \label {RMSE}
\text{RMSE}_{\varphi}^\zeta(\tau) =  \frac{\left\| \mathbf{F'}^{\zeta} - \mathbf{P'}_{\varphi}^{\zeta}\mathbf{R}^{\tau-1} \right\|_2} {\sqrt{{S \times 3L} }},
\end{equation} 
where $\text{RMSE}_{\varphi}^\zeta(\tau)$ represents the RMSE value at time step $\tau$, where \( \tau \in \{1, 2, \ldots, 2L\} \), \( \left\|\cdot\right\|_2\) denotes the matrix 2-norm.
Then, find the minimum RMSE value from all time steps as the similarity value, as follows:
\begin{equation} \label {RMSEmin}
\text{RMSE}_{\varphi}^{*\zeta} = \min_{\tau \in \{1,2,\ldots,2L\}} \text{RMSE}_{\varphi}^\zeta(\tau),
\end{equation} 
where $\text{RMSE}_{\varphi}^{*\zeta}$ is the minimum RMSE value obtained when sliding the particle $\mathbf{P}_{\varphi}^\zeta$ over
the experimental data $\mathbf{F}^{\zeta}$. It represents the similarity between the particle $\mathbf{P}_{\varphi}^\zeta$ and the experimental data $\mathbf{F}^{\zeta}$. A smaller RMSE value indicates a higher similarity. Then, we identify the particle in the set that is most similar to the experimental data, as follows:
\begin{equation} \label {initalParticleindex}
n^{*\zeta} = \arg\min_{ n \in \{1,2,\ldots,N\} } \text{RMSE}_{\varphi}^{*\zeta},
\end{equation}
where $n^{*\zeta}$ is the index of the particle in the set $\mathcal{P}^\zeta$ that is most similar to the experimental data. We retrieve the parameter set corresponding to index $n^{*\zeta}$, and iterate over this parameter set to optimize the particle. Given \(\varphi = (n^{*\zeta}, m) \), the parameters in $\Theta_{\varphi}^\zeta$ updates are as follows:
\begin{equation} \label {Thetaupdate}
    \theta_{\varphi}^\zeta \sim \mathcal{N}\left( \theta_{\varphi-1}^\zeta, (\sigma_{\text{ite}})^2\right),  \quad m = 1, 2, \ldots, M,
\end{equation}
where \( \theta_{\varphi}^\zeta \in \{ A_{\varphi,s}^\zeta, \mu_{\varphi,s}^{\text{ver},\zeta}, \Delta \mu_{\varphi,s}^{\text{hor},\zeta}, \sigma_{\varphi,s}^{\text{ver},\zeta}, \sigma_{\varphi,s}^{\text{hor},\zeta} \mid s = 1, 2, \ldots, S ; \Delta \mu_{\varphi,1}^{\text{hor},\zeta} \text{ is excluded} \} \) is the expansion of the paremeters in $\Theta_{\varphi}^\zeta$, \(\mathcal{N}(\cdot, \cdot)\) represents the normal distribution, \(\varphi_{-1} = (n^{*\zeta}, m-1)\) represents the index of the previous iteration set, $m$ is the iteration number, \( M = 2000 \) is the maximum number of iterations, $\sigma_\text{ite}$ is of the normal distribution governing the iterative update. The value of $\sigma_{\text{ite}}$ depends on the parameter being updated as follows:
\begin{equation} \label {sigmaIte}
    \sigma_{\text{ite}} = 
    \begin{cases}
        0, & \text{for } \theta_{\varphi}^\zeta = \mu_{\varphi,s}^{\text{ver},\zeta}, \\[1em]
        0.1, & \text{for } \theta_{\varphi}^\zeta \in \{ A_{\varphi,s}^\zeta, \sigma_{\varphi,s}^{\text{ver},\zeta} \}, \\[1em]
        1, & \text{for } \theta_{\varphi}^\zeta \in \{ \Delta \mu_{\varphi,s}^{\text{hor},\zeta}, \sigma_{\varphi,s}^{\text{hor},\zeta} \},
    \end{cases}
\end{equation}
where, for the parameter $\mu_{\varphi,s}^{\text{ver},\zeta}$, \(\sigma_{\text{ite}} = 0\) because $\mu_{\varphi,s}^{\text{ver},\zeta}$ is a fixed value. When the parameter is a floating-point number, \(\sigma_{\text{ite}} = 0.1\), and when the parameter is an integer, \(\sigma_{\text{ite}} = 1\).

After each iteration, $\Theta_{\varphi}^\zeta$ generates an updated particle $\mathbf{P}_{\varphi}^\zeta$ using Equations (\ref{generateP}) to (\ref{G}). Then, Equations (\ref{P'}) to (\ref{RMSEmin}) are used to calculate its $\text{RMSE}_{\varphi}^{*\zeta}$. If the value of $\text{RMSE}_{\varphi}^{*\zeta}$ decreases after iterating, $\Theta_{\varphi}^\zeta$ is updated; otherwise, it is not updated, as follows:
\begin{equation} \label {updateTheta}
 \Theta_{\varphi}^\zeta = 
\begin{cases} 
 \Theta_{\varphi}^\zeta, & \text{if } \text{RMSE}_{\varphi}^{*\zeta} < \text{RMSE}_{\varphi_{-1}}^{*\zeta}, \\[1em]
 \Theta_{\varphi_{-1}}^\zeta, & \text{otherwise}.
\end{cases}
\end{equation} 
We obtain the optimal particle parameter set after $M$ iterations, as follows:
\begin{equation} \label {Theta*}
\Theta^{*\zeta} = \Theta_{\varphi}^\zeta, \quad \text{where } \varphi = (n^{*\zeta}, M),
\end{equation}
where \(\Theta^{*\zeta}\) is the optimal parameters for \(\mathbf{F}^\zeta\), meaning that a particle generated using \(\Theta^{*\zeta}\) is most similar to \(\mathbf{F}^\zeta\). Each optimal parameter in the set is as follows:
\begin{align} \label {Theta*parameters}
\Theta^{*\zeta} = \left\{ A_s^{*\zeta} ,\boldsymbol{\mu}_{s}^{*\zeta}, \boldsymbol{\Sigma}_{s}^{*\zeta} \right\},
\end{align} 
where $A_s^{*\zeta}$, $\boldsymbol{\mu}_{s}^{*\zeta}$, and $\boldsymbol{\Sigma}_{s}^{*\zeta}$ represent the optimal $A_s^{\zeta}$, $\boldsymbol{\mu}_{s}^{\zeta}$, and $\boldsymbol{\Sigma}_{s}^{\zeta}$ respectively. Through the above process, for each experimental data $\mathbf{F}^\zeta$ in the set $\mathcal{F}$, we have generated a corresponding optimal parameter set $\Theta^{*\zeta}$. Next, we will take the average of each parameter in the parameter set for the same nodule size \( b \), to represent the characteristics of the data for each nodule size, as follows:
\begin{equation} \label{barTheta}
      \bar{\Theta}^{*b} = \left\{ \langle A_s^{*\zeta} \rangle^b , \langle \boldsymbol{\mu}_{s}^{*\zeta} \rangle^b, \langle \boldsymbol{\Sigma}_{s}^{*\zeta} \rangle^b \right\},
\end{equation}
where $ \bar{\Theta}^{*b} $ represents the averaged parameter set for nodule size $ b $ data, $\langle \cdot \rangle^b$ represents the averaging process for a parameter over all $q = 1, 2, \ldots, Q$ for a given nodule size $b$ is defined as:
\begin{equation} \label{averagingProcess}
\langle X \rangle^b = \frac{1}{Q} \sum_{q=1}^{Q} X.
\end{equation}
By substituting the parameters from the $\bar{\Theta}^{*b}$ set into equations (10) and (11), we can obtain the template particle $\mathbf{T}^b$ for nodule size $b$. For the case without nodules, as shown in Fig.\ref{preprocessing}(A)(3), since it has no pattern, we use a zero matrix as the template for no nodule $\mathbf{T}^0$, as follows:
\begin{equation} \label{T0}
\mathbf{T}^0 = 
\begin{bmatrix}
0_{S \times L}
\end{bmatrix}.
\end{equation}
Fig.\ref{general_template} displays aNll the templates ($\mathbf{T}^0$ to $\mathbf{T}^5$).

\subsection{Nodule Detection Method}
The method for nodule detection involves pre-processing the data collected by sensors and comparing it with the templates described above. The template that best matches the test data is identified as the detection result. The method for comparing the similarity between test data and templates remains the calculation of the minimum RMSE value when the template slides over the test data. This process is the same as the comparison between $\mathbf{F}^\zeta$ and $\mathbf{P}_{\varphi}^\zeta$ mentioned earlier. The calculation of the minimum RMSE value can be referenced from Equations (\ref{F'}) to (\ref{RMSEmin}).

\section{Experimental Results}
\label{results}
We collected $20$ sets of data for cases without nodules and nodules of sizes ranging from $1$ mm to $5$ mm, respectively, for testing. The accuracy of sensors under different nodule sizes and their ability to assess nodule sizes were evaluated. 

\subsection{Precision and Recall}

\begin{figure}[tb]
\centering
\vspace{1mm}
\includegraphics[width=0.7\columnwidth]{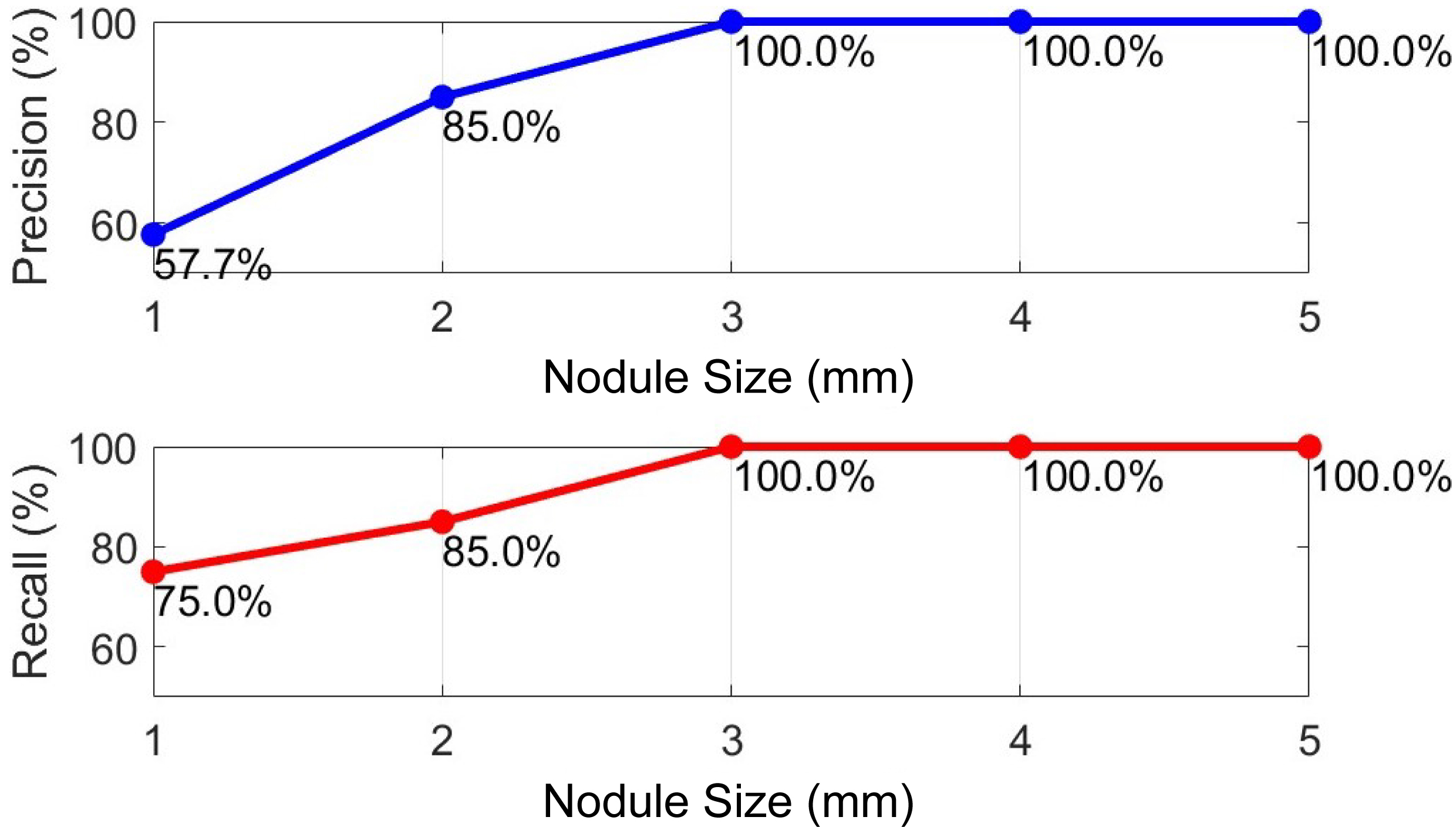}
\caption{Precision and recall for nodule detection. The precision and recall for $1$ mm nodules are relatively low. For $2$ mm nodules, precision and recall have reached $85\%$. For $3$ mm to $5$ mm nodules, precision and recall have reached $100\%$. This indicates that the sensor can effectively detect nodules of $2$ mm or larger.}
\label{precision_recall}
\end{figure}

\begin{figure}[tb]
\centering
\vspace{1mm}
\includegraphics[width=1\columnwidth]{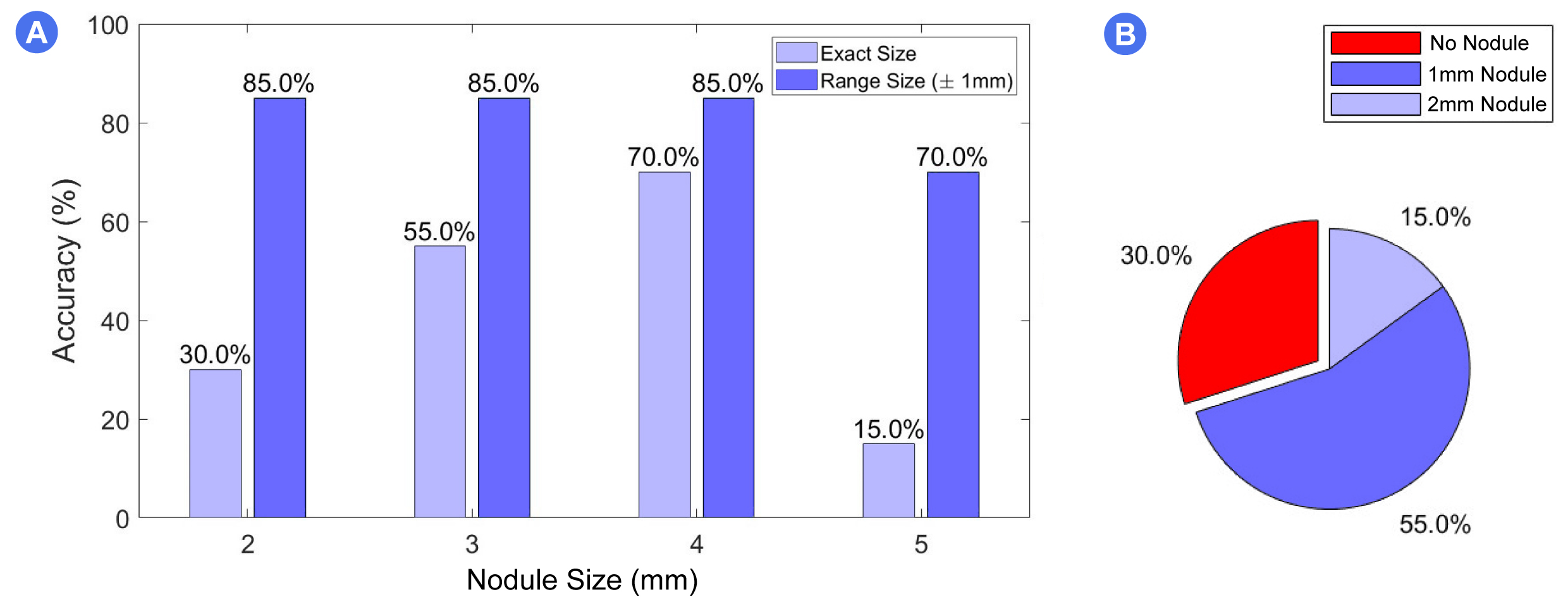}
\caption{(A) Accuracy of nodule size estimation. This indicates that the sensor can provide relatively accurate nodule size estimates within a range of $\pm1$ mm. (B) The distribution of negative results. This indicates that the detection results for $1$ mm nodules are unreliable. }
\label{bar_and_pie}
\end{figure}

We tested whether the sensors could detect nodules under different nodule sizes. The reliability of detection is measured using precision and recall. The definitions of precision and recall are as follows \cite{precisionRecall}:

\begin{equation}
    \text{Precision} = \frac{\text{TP}}{\text{TP} + \text{FP}}, \quad
    \text{Recall} = \frac{\text{TP}}{\text{TP} + \text{FN}},
\end{equation}

where TP is true positive, FP is false positive, and FN is false negative. In this study, precision is the probability that the detection result of the sensor is a nodule and there actually is a nodule. Recall is the probability that the sensor detects a nodule when a nodule actually exists. The experimental results, as shown in Fig.\ref{precision_recall}, reveal that for $1$ mm nodules, the precision is $57.7\%$, indicating that $42.3\%$ of the results detecting $1$ mm nodules actually do not have nodules. The recall rate for $1$ mm nodules is $75\%$, meaning that $25\%$ of $1$ mm nodules are not detected. For $2$ mm nodules, both precision and recall reach $85\%$, indicating that only $15\%$ of the data detecting $2$ mm nodules actually do not have nodules, and $15\%$ of $2$ mm nodules are not detected. For nodules larger than $2$ mm, both precision and recall reach $100\%$, indicating that the sensor can effectively detect nodules of $2$ mm and above. Moreover, the detection performance is best for nodules of $3$ mm or larger. For all negative results (as shown in Fig.\ref{bar_and_pie}(B), $55\%$ are actually $1$ mm nodules, indicating that $1$ mm nodules cannot be detected by this sensor. $2$ mm is the minimum nodule detection size limit for the sensor.

\subsection{The Accuracy of Nodule Size Estimation}


Since the sensor can detect nodules of $2$ mm and above, we analyzed the accuracy of nodule size estimation for $2$ mm to $5$ mm nodules. The results are shown in Fig.\ref{bar_and_pie}(A). For accurate size detection, the success rate of estimating $2$ mm nodules is $30\%$, $3$ mm nodules is $55\%$, $4$ mm nodules reach $70\%$, and $5$ mm nodules only have a success rate of $15\%$. The accuracy increases gradually with nodule size, reaching the highest at $4$ mm, and then decreases for $5$ mm nodules. This is because the similarity between $5$ mm and $4$ mm data is high, and many $5$ mm data are identified as $4$ mm data. For a given tolerance of $\pm1$ mm in size estimation, for instance, a $2$ mm nodule can be estimated as a nodule ranging from $1$ mm to $3$ mm, and so on. Due to the lack of $6$ mm nodule data, the assessment range for $5$ mm nodules is from $4$ mm to $5$ mm, rather than $4$ mm to $6$ mm. The success rate of estimating nodules from $2$ mm to $4$ mm is $85\%$. For $5$ mm nodules, it is $70\%$, possibly because there is no $6$ mm data, making the tolerance smaller than for other nodule sizes. The experimental results indicate that while the sensor struggles to provide precise estimates of nodule sizes, it can effectively provide an approximate range for the nodules, with a tolerance of $\pm1$ mm.

\section{Conclusion}
\label{conclusions}

We proposed a novel 2D origami capsule designed using flexible piezoresistive material to detect early-stage small intestinal nodules. The capsule contained four tactile sensors to measure tissue interaction forces while moving along the intestine. Experiments were conducted using a rubber intestine phantom with peristalsis. A particle filter was designed to estimate the size of nodules in the intestine phantom. The capsule is designed to passively move forward or backward in the intestine only by changing the shape of its origami structure, allowing it to reach any position in the intestine. The proposed method demonstrates reliable detection for nodules of $2$ mm and above, while also assessing the size range of the nodules.


\bibliographystyle{IEEEtran}
\bibliography{References}

\vfill

\end{document}